\documentclass{article}
\usepackage{colm2024_conference}

\usepackage{microtype}
\usepackage{hyperref}
\usepackage{url}
\usepackage{booktabs}
\definecolor{darkblue}{rgb}{0, 0, 0.5}
\hypersetup{colorlinks=true, citecolor=darkblue, linkcolor=darkblue, urlcolor=darkblue}
\usepackage{multirow}
\usepackage{graphicx} 
\usepackage{caption} 
\usepackage{subcaption} 
\usepackage{amsmath}
\usepackage[flushleft]{threeparttable}
\usepackage{colortbl}
\definecolor{lightorange}{HTML}{F59F05}
\definecolor{deeporange}{HTML}{EA5110}

\title{BM$\mathcal{X}$: Entropy-weighted Similarity and Semantic-enhanced Lexical Search \thanks{\ Paying tribute to BM25, the naming style of BM$\mathcal{X}$ follows that of BM25. $\mathcal{X}$ means extension.}}

\hypersetup{colorlinks=true, urlcolor=black}
\author{
    \begin{tabular}[t]{c}
        \textbf{Xianming Li}\textsuperscript{\rm 1}$^\dagger$ \quad
        \textbf{Julius Lipp}\textsuperscript{\rm 2}$^\dagger$ \quad
        \textbf{Aamir Shakir}\textsuperscript{\rm 2} \quad
        \textbf{Rui Huang}\textsuperscript{\rm 2} \quad
        \textbf{Jing Li}\textsuperscript{\rm 1} \\[1ex]
        \midrule
        \multicolumn{1}{c}{\rm\scriptsize $\dagger$ Equal Contribution} \\[1ex]
        \textsuperscript{1} \rm The Hong Kong Polytechnic University, Hong Kong SAR \\
        \textsuperscript{2} \rm Mixedbread, Berlin, Germany \\[1ex]
        \ttfamily
        \rm
        \small
        \href{mailto:xianming.li@connect.polyu.hk}{ xianming.li@connect.polyu.hk} \quad \href{mailto:jing-amelia.li@polyu.edu.hk}{jing-amelia.li@polyu.edu.hk} \\
        \rm
        \small
        \quad \href{mailto:julius@mixedbread.ai}{julius@mixedbread.ai} \quad
        \href{mailto:aamir@mixedbread.ai}{aamir@mixedbread.ai} \\
        \rm
        \small
        \href{mailto:rui@mixedbread.ai}{rui@mixedbread.ai}
    \end{tabular}
}

\colmfinalcopy

\begin{document}

\begin{center}
\maketitle

\begin{abstract}

BM25, a widely-used lexical search algorithm, remains crucial in information retrieval despite the rise of pre-trained and large language models (PLMs/LLMs). However, it neglects query-document similarity and lacks semantic understanding, limiting its performance. We revisit BM25 and introduce BM$\mathcal{X}$, a novel extension of BM25 incorporating entropy-weighted similarity and semantic enhancement techniques. Extensive experiments demonstrate that BM$\mathcal{X}$ consistently outperforms traditional BM25 and surpasses PLM/LLM-based dense retrieval in long-context and real-world retrieval benchmarks. This study bridges the gap between classical lexical search and modern semantic approaches, offering a promising direction for future information retrieval research. The reference implementation of BM$\mathcal{X}$ can be found in Baguetter, which was created in the context of this work. The code can be found here: \url{https://github.com/mixedbread-ai/baguetter}.
\end{abstract}
\end{center}

\section{Introduction}
\label{sec::introduction}

Information retrieval (IR) remains a fundamental need in human daily life \citep{zhu2023large} and has found a renewed significance in the era of large language models (LLMs), particularly in retrieval-augmented generation (RAG) \citep{gao2023retrieval}. Among the various IR techniques, BM25 \citep{bm25} stands out as one of the most influential lexical search algorithms. It is used at the core of many popular search systems such as Lucene, Elasticsearch, or Tantivy. 

Despite the growing focus on text embedding-based semantic search \citep{reimers2019sentence, gao2021simcse, li2023angle, li-li-2024-bellm}, BM25 maintains its widespread use in IR due to its high efficiency and strong generalization ability. The BM25 score function is defined as:
\begin{equation}
    \mathrm{score}(D, Q) = \sum _{i=1}^{m} \mathrm{IDF}(q_i) \cdot \frac{
        \mathrm{F}(q_i, D) \cdot (k_1 + 1)
    }{
        \mathrm{F}(q_i, D) + k_1 \cdot (1 - b + b \cdot \frac{\left | D \right|}{avgdl})
    },
\end{equation}
where $D$, $Q$, $\mathrm{IDF}(\cdot)$, and $\mathrm{F}(\cdot)$ denote the document, the query, inverse document frequency (IDF), and term frequency (TF), respectively. $q_i$ is the $i$-th query token of $Q$.
The algorithm operates by first retrieving documents using each query token via an inverted index, then ranking the retrieved documents using the score function.

BM25 has two significant limitations: (1) It does not consider the similarity between the query and the document, which could enable a more precise assessment of relevance. \\(2) As a lexical search algorithm, it lacks semantic understanding, failing to capture linguistic nuances such as synonyms and homonyms. This limitation is a key factor in the under-performance of lexical search compared to domain-specific text embedding semantic search.

To tackle these challenges, we propose BM$\mathcal{X}$, a lexical search algorithm that incorporates both similarity and semantics. It comes with three key innovations: (1) Entropy-weighted similarity: We utilize the entropy of each query token to weight the similarity scores between the related documents for that token and the query itself. This mitigates bias from high-frequency tokens, ensuring that less frequent but more informative tokens carry higher similarity weight. 
(2) Weighted query augmentation (WQA): To introduce semantics into lexical search, we propose a technique that simultaneously processes the original query and corresponding augmented queries within BM$\mathcal{X}$. This also eliminates the need for multiple retrievals and reranking steps, thus resulting in better efficiency. (3) BM25/BM$\mathcal{X}$ score normalization: To improve the algorithm's applicability in certain scenarios and benefit threshold setting, we normalize their output scores using an estimated maximum value.

We evaluate BM$\mathcal{X}$ against BM25 and several popular PLM and LLM-based text embedding models using widely recognized IR benchmarks, including BEIR \citep{thakur2021beir}, LoCo \citep{jon2024loco}, and BRIGHT \citep{su2024bright}. Our experimental results demonstrate the effectiveness and robustness of BM$\mathcal{X}$, consistently outperforming BM25 across all benchmarks. 
Notably, BM$\mathcal{X}$ surpasses 7 billion parameter LLM text embedding model on the long-context retrieval benchmark LoCo \citep{jon2024loco} and outperforms popular proprietary  embedding models on the BRIGHT benchmark \citep{su2024bright}.

In summary, our contributions are as follows:
\begin{itemize}
    \item We revisit the BM25 algorithm and propose BM$\mathcal{X}$, a lexical search algorithm that addresses the key limitations of its predecessor.

    \item We introduce an entropy-weighted similarity measure into lexical search to retrieve more relevant documents for a given query and design an efficient weighted query augmentation technique to incorporate semantics into lexical search for improved information retrieval. Additionally, we propose a way of normalizing BM25 scores.

    \item We demonstrate through extensive experimental results that BM$\mathcal{X}$ is effective and consistently outperforms BM25 across various information retrieval benchmarks, even rivaling more complex embedding-based approaches in certain scenarios.

    \item For convenient use of BM$\mathcal{X}$, we open-source a flexible, efficient, and hackable search engine library -- \textit{Baguetter}.
\end{itemize}

\section{Related Work}
\label{sec::related_work}

This work aligns with the paradigm of lexical search. Lexical search remains one of the most widely adopted information retrieval (IR) approaches, known for its efficiency, interpretability, and generalization capability.

\paragraph{Lexical Search}

In lexical search, TF-IDF (Term Frequency-Inverse Document Frequency) serves as a fundamental tool for scoring and ranking document relevance given a query \citep{tfidf}. BM25 \citep{bm25} enhances TF-IDF with carefully designed TF and IDF terms. BM25+ \citep{lv2011bm25plus} refines BM25 by introducing a sufficiently large lower bound for TF normalization.

Existing lexical search algorithms typically use query tokens independently to retrieve relevant documents and subsequently rank them based on a score function. These approaches often overlook the overall semantic similarity between the query and the documents, which is crucial in assessing the degree of matching between the query and each individual document. 
To address this limitation, BM$\mathcal{X}$ incorporates similarity techniques to boost the relevance of retrieved documents for a given query, thus improving the search quality.

\paragraph{Query Augmentation}

Query augmentation is an important IR technique for improving the performance of IR systems \citep{carpineto2012survey}. Traditional approaches primarily rely on lexical ontologies, such as WordNet \citep{fellbaum2010wordnet} and FrameNet \citep{baker1998berkeley, li2024comprehensive}, for query augmentation. Yet, these manually annotated lexical resources have limitations in terms of data coverage and currency, which affects their effectiveness in practical applications.

Recently, with the surge of LLMs, there has been growing interest in leveraging these models for query augmentation to improve dense retrieval \citep{baek2024crafting}. Aligning with this trend, we propose the use of LLMs for query augmentation in lexical search. Our approach introduces a one-time query augmentation technique for lexical search, where the retrieval and reranking stages are performed only once, resulting in an augmented retrieval outcome without the need for multiple retrieval cycles. This method aims to enhance efficiency while maintaining the benefits of augmented queries.

\section{Methodology}
\label{sec::methodology}

In this section, we present a detailed exposition of the proposed BM$\mathcal{X}$ algorithm.
We begin in Section \ref{sec::bmx-core} by describing the core algorithm, with particular emphasis on the entropy-weighted similarity mechanism.
Afterward, Section \ref{sec::wqa} introduces weighted query augmentation (WQA), which substantially enhances the semantic capabilities of BM$\mathcal{X}$.
We then describe the normalization process applied to BM$\mathcal{X}$ scores in Section \ref{sec::normalized-bmx}.
To demonstrate the broader applicability of our approaches, we extend score normalization and WQA to the widely-used BM25 algorithm in the appendix section \ref{sec::enhanced-bm25}.

\subsection{Core algorithm of BM$\mathcal{X}$}
\label{sec::bmx-core}
Given a set of documents $\mathcal{D}$ consisting of $\{D_1, D_2, \ldots, D_n\}$ documents where $n$ indicates the total number of documents, and a query $Q$ consisting of $\{q_1, q_2, \ldots, q_m\}$ tokens where $m$ is the total number of query tokens, the BM$\mathcal{X}$ score of a document $D$ is calculated as follows:
\begin{equation}
    \label{eq::bmx-core}
    \mathrm{score}(D, Q) = \sum_i^m \mathrm{IDF}(q_i) \cdot \frac{
            \mathrm{F}(q_i, D) \cdot (\alpha + 1.0)
        }{
            \mathrm{F}(q_i, D) +  \alpha \cdot \frac{\left | D \right |}{avgdl} + \alpha \cdot \mathcal{E}
        } + \beta \cdot \mathrm{E}(q_i) \cdot  \mathrm{S}(Q, D)
\end{equation}
$\left | D \right |$ is the length of the document $D$ in tokens, and $avgdl$ is the average document length in the text collection from which documents are drawn. $\alpha$ and $\beta$ are two hyperparameters. For English IR tasks \footnote{For tasks involving other languages, it is advisable to evaluate and potentially adjust the algorithm's hyperparameters to optimize performance.}, we derive the optimal default parameters in the following way:
\begin{equation}
    \begin{aligned}
    \alpha &= \max\left(\min\left(1.5, \frac{avgdl}{100}\right), 0.5\right) \quad \text{and} \quad
    \beta &= \frac{1}{\log(1 + n)}
    \end{aligned}
\end{equation}
$\mathrm{F}(q_i, D)$ is the number of times that the token $q_i$ occurs in the document $D$.
$\mathrm{IDF}(q_i)$ represents the IDF (inverse document frequency) weight of the query token $q_i$. 
Its calculation is the same as BM25 \citep{bm25}, as follows:
\begin{equation}
    \begin{split}
       \mathrm{IDF}(q_i) = \mathrm{log}\left (\frac{n - l + 0.5}{l + 0.5} + 1 \right ),
    \end{split}
\end{equation}  
where $l$ is the number of documents containing the query token $q_i$, and the corresponding document subset containing $q_i$ is $\mathcal{D}^s \subseteq \mathcal{D}$ consisting of $\{D^s_1, D^s_2, \ldots, D^s_l\}$.

$\mathrm{E}(q_i)$ is the normalized entropy of the query token $q_i$ and is expressed as:
\begin{equation}
    \begin{split}
        \mathrm{E}(q_i) &= \frac{
            \mathrm{\Tilde{E}}(q_i)
        }{
            \mathrm{max}(\mathrm{\Tilde{E}}(q_1), ..., \mathrm{\Tilde{E}}(q_m))
        } \\ \\
        \text{where} \quad \mathrm{\Tilde{E}}(q_i) &= - \sum_{j}^{l} p_j\ \mathrm{log} p_j \\
        \quad \ p_j &= \mathrm{sigmoid}(\mathrm{F}(q_i, D^s_j)) \\
        &= \frac{1}{1 + \mathrm{exp}(-\mathrm{F}(q_i, D^s_j))}
    \end{split}
\end{equation}
$\mathrm{F}(q_i, D^s_j)$ is the number of times that the token $q_i$ occurs in the document $D^s_j$. $\mathcal{E}$ is the average entropy for query $Q$, as follows:
\begin{equation}
    \mathcal{E} = \frac{\sum_i^m \mathrm{E}(q_i)}{m}.
\end{equation}
$\mathrm{S}(Q, D)$ is the similarity between the query $Q$ and document $D$. There are various similarity measurements that can be used. Here we adopt a simple one, as follows:
\begin{equation}
    \mathrm{S}(Q, D) = \frac{ \left | Q \cap  D \right| }{ m },
\end{equation}
where $\left | Q \cap  D \right|$ denotes the number of common tokens between query and document.

In BM$\mathcal{X}$, we address the impact of high-frequency tokens, particularly in documents with numerous repetitions, to maintain balanced probability distributions. We employ the sigmoid function to derive probability values from token frequencies, effectively mapping frequencies exceeding five to similar values. This approach imposes an implicit cap on the maximum effective token frequency, increasing the algorithm's robustness against distortions caused by highly repetitive text. By doing so, BM$\mathcal{X}$ ensures relevant contribution from both, maintaining effective retrieval performance across different document structures and content types. 
By taking the entropy-weighted similarity into account, BM$\mathcal{X}$ can retrieve more relevant documents compared to BM25.

\subsection{WQA: Weighted Query Augmentation}
\label{sec::wqa}
Lexical search, while efficient, lacks the semantic understanding of text embedding-based semantic search techniques \citep{li2023angle}. It retrieves documents based solely on query tokens, overlooking, e.g., synonyms and homonyms. To address this limitation, we propose query augmentation using Large Language Models (LLMs):
\begin{equation}
\mathcal{Q}^A = \mathrm{LLM}(Q, t),
\end{equation}
where $t$ is the number of augmented queries, and $\mathcal{Q}^A$ represents the set of augmented queries $\{Q_1^A, Q_2^A, \ldots, Q_t^A\}$.
Rather than performing multiple retrieval iterations with the original and each augmented query, we introduce a weighted query augmentation method. This approach combines the original and augmented queries into a single weighted query, eliminating the need for multiple queries:
\begin{equation}
\mathrm{score}(D, Q, \mathcal{Q}^A) = \mathrm{score}(D, Q) + \sum_{i=1}^t w_i \cdot \mathrm{score}(D, Q_i^A),
\end{equation}
where $w_i$, typically within $[0,1]$, stands for the weight assigned to the $i$-th augmented query and $t$ for the total number of augmented queries. This weighting scheme introduces semantic understanding into lexical search while preserving computational efficiency.

\subsection{Normalized BM$\mathcal{X}$}
\label{sec::normalized-bmx}
The BM$\mathcal{X}$ score from equation (\ref{eq::bmx-core}) is unnormalized, which may impact its application in certain scenarios, such as threshold-based retrieval. To address this, we propose a normalized version of BM$\mathcal{X}$ using maximum normalization. We begin by estimating the maximum possible score as follows:
\begin{equation}
    \mathrm{score}(D, Q)^{max} = \sum_i^m \mathrm{max}(\mathrm{IDF}(q_i)) + 1.0 = m \cdot \left [ \mathrm{log} \left (1 + \frac{ n - 0.5}{1.5} \right ) + 1.0 \right ]
\end{equation}
To calculate the normalized BM$\mathcal{X}$ score, we then apply:
\begin{equation}
\mathrm{score}(D, Q)^{norm} = \frac{\mathrm{score}(D, Q)}{\mathrm{score}(D, Q)^{max}}
\end{equation}
This normalization keeps BM$\mathcal{X}$ scores within interval $[0, 1]$, benefiting threshold retrieval.

\section{Experiment}
\label{sec::experiment}

In the following experiment results, BM$\mathcal{X}$ refers to the core algorithm, while BM$\mathcal{X}$ + WQA indicates BM$\mathcal{X}$ with weighted query augmentation.

\subsection{Experiment Setup}

\label{sec::setup}
\paragraph{Dataset.}
To comprehensively evaluate the proposed BM$\mathcal{X}$, we conduct experiments on various popular IR benchmarks, including BEIR \citep{thakur2021beir}, long-text retrieval LoCo \citep{jon2024loco}, and the more realistic and challenging BRIGHT \citep{su2024bright} benchmark. 
For convenient comparison with embedding-based models, we adopt the MTEB \citep{muennighoff2022mteb} variant for BEIR.

\paragraph{Implementation.} We implemented BM$\mathcal{X}$ using \textit{Baguetter}, an open-source Python library developed in the context of this research. \textit{Baguetter} is a significantly modified fork of retriv \citep{retriv2023} and provides a unified search interface across sparse, dense, and hybrid retrieval methods. It incorporates an efficient BM25 implementation from the BM25S project \citep{bm25s2024}, which we abstracted to simplify comparisons with BM$\mathcal{X}$. The library supports multi-threaded indexing and searching, making it suitable for large-scale experiments. Its flexible architecture allows for customizable text processing pipelines, custom BM25 algorithms, simple retrieval evaluation, embedding quantization experiments and more.
\textit{Baguetter}'s design philosophy has simplicity at its core, enabling rapid prototyping and efficient iteration on algorithm and model development. This approach sped up our experimental process while maintaining necessary flexibility. The complete source code, including our BM$\mathcal{X}$ implementation and evaluation scripts, is available at:\\
\url{https://github.com/mixedbread-ai/baguetter}.

\subsection{BEIR Benchmark Results}
\label{sec::beir}
For the BEIR benchmark, we evaluate on the following datasets: ArguAna (AA), SciDOCS (SD), SciFact (SF), NFCorpus (NFC), TRECCOVID (TCV), Touche2020 (TCH), FiQA2018 (FQA), HotpotQA (HQA), MSMARCO-dev (MM), Fever (FVR), DBPedia (DBP), Quora (QRA), CQADupstack (CQA), and Climate-Fever (CF).
For baselines, we choose widely-used embedding models: BGE (\textit{BAAI/bge-large-en-v1.5}) \citep{bge_embedding},  and Mixedbread (\textit{mixedbread-ai/mxbai-embed-large-v1}) \citep{emb2024mxbai}. For lexical models, we select BM25 with different variants: ATIRE \citep{trotman2014improvements}, Robertson \citep{bm25}, BM25+ \citep{lv2011bm25plus}, BM25L \citep{lv2011bm25plus}, and default kernel Lucene. \footnote{Lucene is an open-source search software: \url{https://lucene.apache.org}} 
We include two versions of BM25: An efficient implementation from the BM25S project \citep{lu2024bm25s} and the implementation in \textit{Baguetter}. To ensure a fair comparison, both BM25 and BM$\mathcal{X}$ utilize an identical text preprocessing pipeline in \textit{Baguetter}.

\begin{table*}[htbp]
\setlength\tabcolsep{3pt}
\scriptsize
\centering
\begin{threeparttable}
\begin{tabular}{lccccccccccccccccc}
\toprule
Model & AA & SD & SF & NFC & TCV & TCH & FQA & HQA & MM & FVR & NQ & DBP & QRA & CF & CQA & Avg. \\

\midrule
\midrule
\multicolumn{17}{c}{\textit{Embedding Models}} \\

\midrule
MiniLM & $50.17$ & $21.64$ & $64.51$ & $31.59$ & $47.25$ & $16.9$ & $36.87$ & $46.51$ & $36.54$ & $51.93$ & $43.87$ & $32.33$ & $87.56$ & $20.27$ & $41.32$ & $41.95$ \\
BGE & $63.61$ & $21.73$ & $74.04$ & $37.39$ & \cellcolor{lightorange}$78.07$ & \cellcolor{lightorange}$25.70$ & $40.65$ & \cellcolor{lightorange}$72.60$ & \cellcolor{lightorange}$41.35$ & $86.29$ & $54.15$ & $40.77$ & $88.90$ & \cellcolor{lightorange}$42.23$  & $36.57$ & $53.60$ \\ 
MXBAI & \cellcolor{lightorange}$66.02$ & \cellcolor{lightorange}$23.32$ & \cellcolor{lightorange}$74.73$ & \cellcolor{lightorange}$38.64$ & $75.57$ & $25.20$ & \cellcolor{lightorange}$45.27$ & $72.03$ & $41.26$ & \cellcolor{lightorange}$86.91$ & \cellcolor{lightorange}$55.79$ & \cellcolor{lightorange}$44.51$ & \cellcolor{lightorange}$88.98$ & $36.09$ & \cellcolor{lightorange}$41.6$ & \cellcolor{lightorange}$54.39$ \\

\midrule
\midrule
\multicolumn{17}{c}{\textit{Lexical Models}} \\
\midrule

\multicolumn{17}{l}{\textit{BM25s \citep{lu2024bm25s}, k1=1.2, b=0.75}} \\
Robertson & $49.2$ & $15.5$ & $68.3$ & $31.9$ & $59.0$ & $33.8$ & $25.4$ & $58.5$ & $22.6$ & $50.3$ & $29.2$ & $30.3$ & $80.4$ & $13.7$ & $29.9$ & $39.87$ \\
ATIRE & $48.7$ & $15.6$ & $68.1$ & $31.8$ & $61.0$ & $33.2$ & $25.3$ & $58.5$ & $22.6$ & $50.3$ & $29.1$ & $30.3$ & $80.5$ & $13.7$ & $30.1$ & $39.92$ \\
BM25+ & $48.7$ & $15.6$ & $68.1$ & $31.8$ & $61.0$ & $33.2$ & $25.3$ & $58.5$ & $22.6$ & $50.3$ & $29.1$ & $30.3$ & $80.5$ & $13.7$ & $30.1$ & $39.92$ \\
BM25L & $49.6$ & $15.8$ & $68.7$ & $32.2$ & $62.9$ & $33.0$ & $25.0$ & $55.9$ & $21.4$ & $46.6$ & $28.1$ & $29.4$ & $80.3$ & $13.5$ & $29.8$ & $39.48$ \\
BM25 & $48.7$ & $15.6$ & $68.0$ & $31.8$ & $61.0$ & $33.2$ & $25.3$ & $58.5$ & $22.6$ & $50.3$ & $29.1$ & $30.3$ & \cellcolor{lightorange} $80.5$ & \cellcolor{lightorange} $13.7$ & $30.1$ & $39.91$ \\
\multicolumn{17}{l}{\textit{Baguetter, k1=1.2, b=0.75}} \\
Robertson & $49.55$ & $15.69$ & $68.71$ & $32.85$ & $64.36$ & $30.95$ & $25.27$ & $59.13$ & $23.19$ & $50.93$ & $28.55$ & $31.69$ & $78.73$ & $13.54$ & $30.73$ & $40.26$ \\
ATIRE & $49.23$ & $15.66$ & $68.73$ & $32.71$ & $65.94$ & $31.15$ & $25.30$ & $59.13$ & $23.23$ & $50.98$ & $28.59$ & $31.56$ & $78.72$ & $13.55$ & $31.00$ & $40.37$ \\
BM25+ & $49.23$ & $15.66$ & $68.73$ & $32.71$ & $65.94$ & $31.15$ & $25.30$ & $59.13$ & $23.23$ & $50.98$ & $28.59$ & $31.56$ & $78.72$ & $13.55$ & \cellcolor{lightorange} $31.00$ & $40.37$ \\
BM25L & $50.32$ & $15.88$ & $69.37$ & \cellcolor{lightorange}$33.00$ & $67.05$ & $31.52$ & $24.79$ & $56.35$ & $22.01$ & $47.37$ & $27.35$ & $30.87$ & $78.21$ & $13.29$ & $30.87$ & $39.88$ \\
BM25 & $49.32$ & $15.65$ & $68.67$ & $32.68$ & $65.78$ & $31.11$ & $25.26$ & $59.09$ & $23.24$ & $50.98$ & $28.85$ & $31.56$ & $78.73$ & $13.55$ & $30.99$ & $40.36$ \\
\midrule
BM$\mathcal{X}$ & \cellcolor{lightorange}$50.46$ & \cellcolor{lightorange}$15.91$ & \cellcolor{lightorange}$69.42$ &$32.84$ & \cellcolor{lightorange}$68.1$ & \cellcolor{lightorange}$34.34$ & \cellcolor{lightorange}$25.39$ & \cellcolor{lightorange}$61.73$ & \cellcolor{lightorange}$24.21$ & \cellcolor{lightorange}$55.75$ & \cellcolor{lightorange}$29.84$ & \cellcolor{lightorange}$32.2$ & $78.56$ & $13.32$ & $30.77$ & \cellcolor{lightorange}$41.52$ \\

\bottomrule
\end{tabular}
\end{threeparttable}
\caption{Results on the BEIR benchmark are presented, with NDCG@10 as the primary metric. Model abbreviations are as follows: MiniLM refers to \textit{sentence-transformers/all-MiniLM-L6-v2}, BGE to \textit{BAAI/bge-large-en-v1.5} \citep{bge_embedding}, and MXBAI to \textit{mixedbread-ai/mxbai-embed-large-v1} \citep{emb2024mxbai}. The best values for embedding and lexical models are highlighted in orange. 
The results from BM25S \citep{lu2024bm25s} specify precision limited to one decimal place.
}
\label{table::beir}
\end{table*}

Table \ref{table::beir} presents the main results on the BEIR benchmark. BM$\mathcal{X}$ generally outperforms all BM25 variants, achieving the best results in 11 out of 15 cases. This suggests the effectiveness of BM$\mathcal{X}$ and underscores the importance of incorporating query-document similarity in IR. Compared to BM25S, our implementation in \textit{Baguetter} shows better performance, indicating its high-quality IR abilities. We observe that embedding-based models consistently outperform lexical models. This is attributed to the powerful semantic understanding of embedding-based models, highlighting the importance of semantics in IR.

\begin{table*}[htbp]
\setlength\tabcolsep{10pt}
\scriptsize
\centering
\begin{threeparttable}
\begin{tabular}{lcccccccc}
\toprule
Model & SF & SD & NFC & TCV & DBP & NQ & FQA & Avg. \\

\midrule
\midrule
\multicolumn{9}{c}{\textit{Embedding Models}} \\
\midrule
all-MiniLM-L6-v2 & $64.51$ & $\mathbf{21.64}$ & $31.59$ & $47.25$ & $32.33$ & $\mathbf{43.87}$ & $\mathbf{36.87}$ & $39.72$ \\

\midrule
\midrule
\multicolumn{9}{c}{\textit{Without Weighted Query Augmentation}} \\
\midrule

BM25 (baguetter) & $68.67$ & $15.65$ & $32.68$ & $65.78$ & $31.56$ & $28.85$ & $25.26$ & $38.34$ \\
BM$\mathcal{X}$ & \cellcolor{lightorange}$69.42$ & \cellcolor{lightorange}$15.91$ & \cellcolor{lightorange}$32.84$ & \cellcolor{lightorange}$68.10$ & \cellcolor{lightorange}$32.20$ & \cellcolor{lightorange}$29.84$ & \cellcolor{lightorange}$25.39$ & \cellcolor{lightorange}$39.10$ \\

\midrule
\midrule
\multicolumn{9}{c}{\textit{With Weighted Query Augmentation}} \\
\midrule

BM25 (baguetter) & \cellcolor{lightorange}$\mathbf{70.92}$ & $16.21$ & $34.99$ & $72.04$ & $33.81$ & $33.02$ & $29.09$ & $41.44$ \\

BM$\mathcal{X}$ & $70.83$ & \cellcolor{lightorange}$16.34$ & \cellcolor{lightorange}$\mathbf{35.08}$ & \cellcolor{lightorange}$\mathbf{72.98}$ & \cellcolor{lightorange}$\mathbf{34.52}$ & \cellcolor{lightorange}$35.11$ & \cellcolor{lightorange}$29.48$ & \cellcolor{lightorange}$\mathbf{42.05}$ \\

\bottomrule
\end{tabular}
\end{threeparttable}
\caption{Results on BEIR benchmark with weighted query augmentation. The augmentation size is 10. NDCG@10 serves as the main metric. all-MiniLM-L6-v2 indicates the embedding model \textit{sentence-transformers/all-MiniLM-L6-v2}. The orange color indicates the best value in each model category. The bold numbers represent the best values across all models. The prompt used in weighted query augmentation is specified in Appendix Section \ref{sec::query-aug-prompt}.
}
\label{table::augment-beir}
\end{table*}

To evaluate the impact of semantics on lexical search, we compare lexical search performance with and without query augmentation, as shown in Table \ref{table::augment-beir}. The results demonstrate that with the proposed weighted query augmentation (WQA), both BM25 and BM$\mathcal{X}$ outperform their non-augmented counterparts.

In general, BM$\mathcal{X}$ consistently outperforms BM25 both with and without query augmentation, further validating its effectiveness. Interestingly, BM$\mathcal{X}$ with WQA even surpassing the performance of the embedding-based model \textit{sentence-transformers/all-MiniLM-L6-v2}.

\subsection{LoCo Benchmark Results}
\label{sec::loco}
For the long-context LoCo retrieval benchmark, we evaluate on Tau Scrolls Gov. Report (GovReport), Tau Scrolls QMSUM (QMSUM), QASPER - Title to Article (Qasper-TA), QASPER - Abstract to Article (Qasper-AA), and Tau Scrolls Summ. Screen (SummScreen).
Table \ref{table::loco} presents the results, indicating that even proprietary embedding models struggle with long-context retrieval. In contrast, lexical models generally outperform embedding-based models in this context.
The proposed BM$\mathcal{X}$ outperforms both BM25 and embedding models in most cases, with the exception of M2-BERT-32768, underlining its strong performance in handling long-context retrieval scenarios.

\begin{table*}[htbp]
\setlength\tabcolsep{9pt}
\scriptsize
\centering
\begin{threeparttable}
\begin{tabular}{lcccccc}
\toprule
Model & GovReport & QMSUM & Qasper-TA & Qasper-AA & SummScreen & Avg. \\

\midrule
\midrule
\multicolumn{7}{c}{\textit{Open-sourced Embedding Models}} \\
\midrule
UAE-Large & $93.84$ & $45.41$ & $89.62$ & $95.64$ & $68.46$ & $78.59$ \\
E5-Mistral-7B $\dagger$ & $98.30$ & $46.80$ & $98.40$ & \cellcolor{lightorange}$\mathbf{99.80}$ & $95.90$ & $87.84$ \\
M2-BERT-2048 $\dagger$ & $94.70$ & $58.50$ & $87.30$ & $95.50$ & $81.80$ & $83.56$ \\
M2-BERT-32768 $\dagger$ & \cellcolor{lightorange}$98.50$ & \cellcolor{lightorange}$\mathbf{69.50}$ & \cellcolor{lightorange}$\mathbf{97.40}$ & $98.70$ & \cellcolor{lightorange}$\mathbf{98.60}$ & \cellcolor{lightorange}$\mathbf{92.54}$ \\
\midrule
\midrule
\multicolumn{7}{c}{\textit{Proprietary Embedding Models}} \\
\midrule

OpenAI ada-002 $\dagger$ & $44.30$ & $7.30$ & $85.10$ & $89.70$ & $37.30$ & $52.74$ \\
Cohere v3 $\dagger$ & \cellcolor{lightorange}$72.4$ & \cellcolor{lightorange}$38.10$ & \cellcolor{lightorange}$85.20$ & \cellcolor{lightorange}$91.10$ & \cellcolor{lightorange}$46.40$ & \cellcolor{lightorange}$66.64$ \\
Voyage v1 $\dagger$ & $39.4$ & $15.2$ & $38.4$ & $32.2$ & $1.95$ & $25.43$ \\

\midrule
\midrule
\multicolumn{7}{c}{\textit{Lexical Models}} \\
\midrule
BM25 (\textit{Baguetter}) & $98.76$ & \cellcolor{lightorange}$59.52$ & $94.65$ & $99.46$ & $97.51$ & $89.98$ \\
BM$\mathcal{X}$ & \cellcolor{lightorange}$\mathbf{98.82}$ & $59.21$ & \cellcolor{lightorange}$95.19$ & \cellcolor{lightorange}$99.49$ & \cellcolor{lightorange}$97.89$ & \cellcolor{lightorange}$90.12$ \\
\bottomrule
\end{tabular}
\end{threeparttable}
\caption{Results of LoCo benchmark. NDCG@10 is the reported metric. $\dagger$ indicates results are retrieved from \citep{jon2024loco}. 
UAE-Large is \textit{WhereIsAI/UAE-Large-V1} \citep{li2023angle}. The orange color indicates the best value in each model category. The bold numbers represent the best values across all models.
}
\label{table::loco}
\end{table*}

\subsection{BRIGHT Benchmark Results}
\label{sec::bright}
\begin{table*}[htbp]
\setlength\tabcolsep{2.5pt}
\scriptsize
\centering
\begin{threeparttable}
\begin{tabular}{lcccccccccccccc}
\toprule
Model & BIO & ES & ECO & PSY & ROB & SO & SL & LC & PONY & AOPS & TQ & TT & Avg. \\

\midrule
\midrule
\multicolumn{14}{c}{\textit{Embedding Models}} \\
\midrule
UAE-Large & $13.24$ & $26.48$ & $16.89$ & $18.54$ & $12.05$ & $10.78$ & $13.79$ & $25.18$ & \cellcolor{lightorange}$6.15$ & $6.03$ & $12.62$ & $5.70$ & $13.95$ \\
MXBAI-Large & $15.30$ & \cellcolor{lightorange}$28.04$ & \cellcolor{lightorange}$\mathbf{18.82}$ & \cellcolor{lightorange}$20.05$ & $13.60$ & \cellcolor{lightorange}$11.90$ & $14.38$ & $25.40$ & $5.30$ & $6.10$ & $13.40$ & $4.97$ & $14.77$ \\
E5-Mistral-7B $\dagger$ & \cellcolor{lightorange}$18.80$ & $26.00$ & $15.50$ & $15.80$ & \cellcolor{lightorange}$16.40$ & $9.80$ & \cellcolor{lightorange}$\mathbf{18.50}$ & \cellcolor{lightorange}$28.70$ & $4.80$ & \cellcolor{lightorange}$7.10$ & \cellcolor{lightorange}$23.90$ & \cellcolor{lightorange}$\mathbf{25.10}$ & \cellcolor{lightorange}$17.50$ \\
\midrule
\midrule
\multicolumn{14}{c}{\textit{Proprietary Models}} \\
\midrule
OpenAI $\dagger$ & \cellcolor{lightorange}$23.70$ & $26.30$ & $20.00$ & \cellcolor{lightorange}$\mathbf{27.50}$ & $12.90$ & $12.50$ & \cellcolor{lightorange} $20.30$ & $23.60$ & \cellcolor{lightorange}$2.50$ & \cellcolor{lightorange}$8.50$ & $22.20$ & $10.80$ & $17.57$ \\
Cohere $\dagger$ & $19.00$ & \cellcolor{lightorange}$27.50$ & \cellcolor{lightorange}$20.20$ & $21.80$ & \cellcolor{lightorange}$16.20$ & \cellcolor{lightorange}$16.50$ & $17.70$ & $26.80$ & $1.80$ & $6.50$ & $15.10$ & $7.10$ & $16.35$ \\
Voyage $\dagger$ & $23.60$ & $25.10$ & $19.80$ & $24.80$ & $11.20$ & $15.00$ & $15.60$ & \cellcolor{lightorange}$\mathbf{30.60}$ & $1.50$ & $7.40$ & \cellcolor{lightorange}$\mathbf{26.10}$ & \cellcolor{lightorange}$11.10$ & \cellcolor{lightorange}$17.65$ \\

\midrule
\midrule
\multicolumn{14}{c}{\textit{Lexical Models}} \\
\midrule
BM25 ($k1=0.9$, $b=0.4$) $\dagger$ & $19.20$ & $27.10$ & $14.90$ & $12.50$ & $13.50$ & $16.50$ & $15.20$ & $24.40$ & $7.90$ & $6.20$ & $9.80$ & \cellcolor{lightorange}$4.80$ & $14.33$ \\
\multicolumn{14}{l}{\textit{Baguetter} $\downarrow$} \\
BM25 ($k1=1.2$, $b=0.75$) & $8.92$ & $12.87$ & $10.39$ & $9.16$ & $7.66$ & $14.09$ & $8.84$ & $23.83$ & $1.45$ & $7.43$ & $7.63$ & $1.31$ & $9.47$ \\
BM25 ($k1=0.9$, $b=0.4$) & $19.25$ & $29.67$ & $16.89$ & $15.77$ & $14.19$ & $15.22$ & $15.30$ & $25.05$ & $4.62$ & $8.28$ & $6.57$ & $2.34$ & $14.43$ \\
BM25 ($k1=0.9$, $b=0.4$) + WQA & $22.9$ & $33.86$ & $17.2$ & $24.33$ & $14.54$ & \cellcolor{lightorange}$\mathbf{20.21}$ & \cellcolor{lightorange}$16.01$ & $26.61$ & $6.03$ & $8.41$ & \cellcolor{lightorange}$10.50$ & $3.29$ & $16.99$ \\
\midrule
BM$\mathcal{X}$ & $12.25$ & $16.54$ & $13.14$ & $11.75$ & $10.96$ & $15.13$ & $10.78$ & $24.04$ & $2.68$ & $6.26$ & $7.10$ & $1.24$ & $10.99$ \\
BM$\mathcal{X}$ ($\alpha=0.05$) & $26.4$ & $33.55$ & $13.29$ & $16.06$ & $14.68$ & $13.80$ & $15.47$ & $25.35$ & \cellcolor{lightorange}$\mathbf{10.91}$ & $9.18$ & $8.17$ & $2.38$ & $15.77$ \\
BM$\mathcal{X}$ ($\alpha=0.05$) + WQA & \cellcolor{lightorange}$\mathbf{31.57}$ & \cellcolor{lightorange}$\mathbf{40.24}$ & \cellcolor{lightorange}$17.40$ & \cellcolor{lightorange}$23.53$ & \cellcolor{lightorange}$\mathbf{16.67}$ & $18.32$ & $15.05$ & \cellcolor{lightorange}$28.34$ & $10.82$ & \cellcolor{lightorange}$\mathbf{9.22}$ & $9.98$ & $2.74$ & \cellcolor{lightorange}$\mathbf{18.66}$ \\
\bottomrule
\end{tabular}
\end{threeparttable}
\caption{Results on the BRIGHT benchmark. NDCG@10 serves as the main metric. $\dagger$ indicates results are retrieved from \citep{su2024bright}. 
UAE-Large is \textit{WhereIsAI/UAE-Large-V1} \citep{li2023angle}. 
MXBAI-Large denotes \textit{mixedbread-ai/mxbai-embed-large-v1} \citep{emb2024mxbai}. The orange color indicates the best value in each model category. The bold numbers represent the best values across all models. The prompt used in WQA is specified in Appendix Section \ref{sec::query-aug-prompt}.
}
\label{table::bright}
\end{table*}

To evaluate the BM$\mathcal{X}$ model's performance in realistic retrieval scenarios, we employ the BRIGHT benchmark, utilizing datasets from diverse domains: biology (BIO), earth science (ES), economics (ECO), psychology (PSY), robotics (ROB), StackOverflow (SO), sustainable living (SL), leetcode (LC), pony (PONY), aops (AOPS), theoremqa questions (TQ), and theoremqa theorems (TT).

Table \ref{table::bright} presents the results, demonstrating that BM$\mathcal{X}$ with weighted query augmentation (WQA) outperforms all baselines, including embedding models and other lexical models. This may be attributed to BM$\mathcal{X}$'s ability to generalize effectively across various domains, while embedding models often struggle with out-of-domain data.

The results highlight that the WQA mechanism successfully enhances BM$\mathcal{X}$'s semantic understanding, enabling it to handle realistic retrieval scenarios more effectively than embedding models. 
BM$\mathcal{X}$ achieves this without the need for costly training on massive high-quality datasets, as required by embedding models.

\subsection{Multilingual Results}
\label{sec::multilingual}
To evaluate BM$\mathcal{X}$'s performance across languages, we conducted experiments on five multilingual retrieval datasets from Hugging Face: MMarcoRetrieval (Chinese) from C-MTEB/MMarcoRetrieval, JaQuAD (Japanese) from SkelterLabsInc/JaQuAD, StrategyQA (Korean) from taeminlee/Ko-StrategyQA, GermanDPR (German) from deepset/germandpr, and FQuAD (French) from manu/fquad2\_test.

Table \ref{table::multilingual} presents the results, demonstrating that BM$\mathcal{X}$ consistently outperforms BM25 across these multilingual datasets. This suggests the effectiveness of BM$\mathcal{X}$ in handling diverse multilingual IR tasks.

\begin{table*}[htbp]
\setlength\tabcolsep{7.5pt}
\small
\centering
\begin{threeparttable}
\begin{tabular}{lcccccc}
\toprule
\multirow{2}{*}{\textbf{Model}} &  MMarcoRetrieval & JaQuAD & StrategyQA  & GermanDPR & FQuAD & \multirow{2}{*}{\textbf{Avg.}} \\

& {\scriptsize Chinese} & {\scriptsize Japanese} & {\scriptsize Korean}  & {\scriptsize German} & {\scriptsize French} \\
\toprule

BM25 & $47.41$ & $54.26$ & $33.89$ & $52.27$ & $91.80$ & $55.93$ \\

BM$\mathcal{X}$ & $\mathbf{47.94}$ & $\mathbf{54.63}$ & $\mathbf{35.72}$ & $\mathbf{53.58}$ & $\mathbf{91.92}$ & $\mathbf{56.76}$ \\

\bottomrule
\end{tabular}
\end{threeparttable}
\caption{Results of multilingual retrieval. The parameters for BM25 are set to $k1=1.2$ and $b=0.75$. The parameters for BM$\mathcal{X}$ are set to $\alpha=0.5$ and $\beta=0.1$. The bold numbers represent the best values across all models. }
\label{table::multilingual}
\end{table*}

\subsection{Efficiency Study}
\label{sec::efficiency}
\begin{figure}
     \centering
     \begin{subfigure}[b]{0.49\textwidth}
         \centering
         \includegraphics[width=\textwidth]{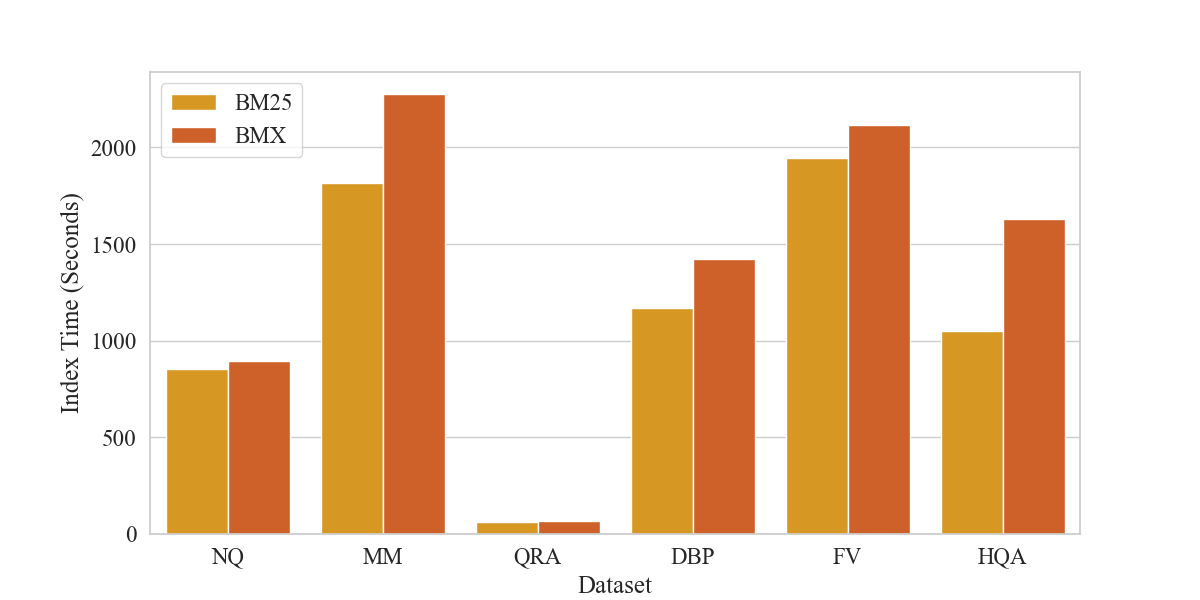}
         \caption{Index Time}
         \label{figure::index-time}
     \end{subfigure}
     \hfill
     \begin{subfigure}[b]{0.49\textwidth}
         \centering
         \includegraphics[width=\textwidth]{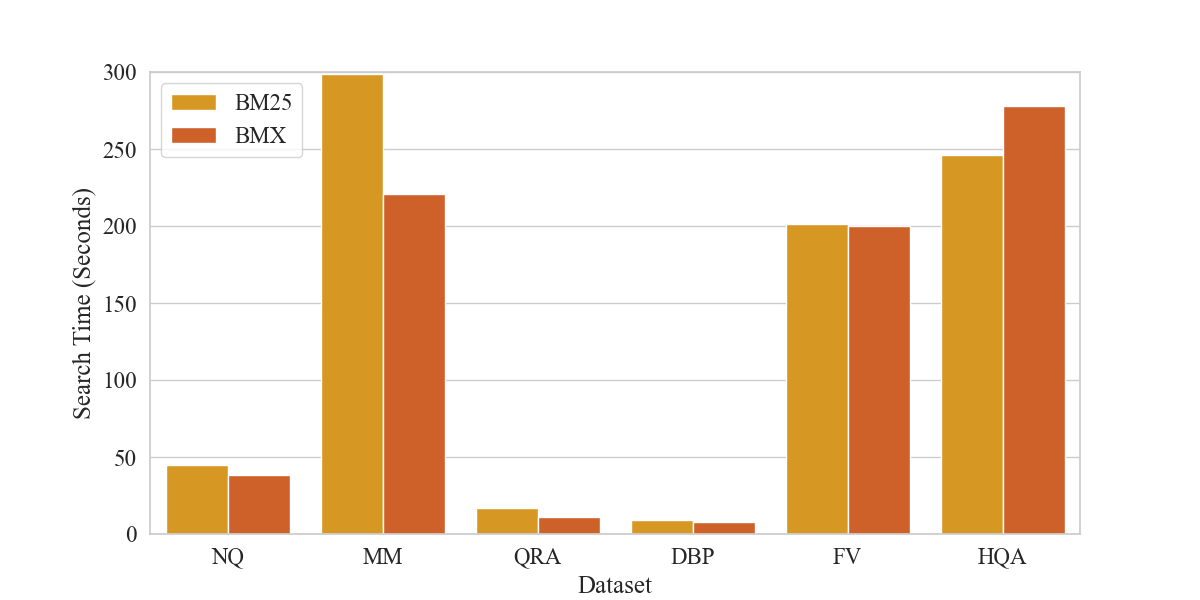}
         \caption{Search Time}
         \label{figure::search-time}
    \end{subfigure}
    \caption{(a) shows the index time of BM25 and BM$\mathcal{X}$ on different retrieval datasets. (b) shows the corresponding search time.}
    \label{figure::efficiency}
\end{figure}
We compare \textit{Baguetter's} BM$\mathcal{X}$ and \textit{Baguetter's} BM25 computational efficiency using BEIR benchmark datasets \citep{thakur2021beir} in a single-threaded environment. Figure \ref{figure::efficiency} shows index and search times. BM25 demonstrates faster indexing due to its simpler computations. However, BM$\mathcal{X}$ sometimes outperforms BM25 in search speed for larger datasets and overall shows a very similar performance.

\subsection{Case Study}
\label{sec::case-study}
Table \ref{table::case} presents a comparative case study of BM25, BM$\mathcal{X}$, and BM$\mathcal{X}$ with WQA. BM25's second retrieved context misses the ``youngest player'' query intent, highlighting its lack of semantic understanding. BM$\mathcal{X}$ captures both ``youngest'' and ``Premier League'' intents in its top 3 results, though the second document refers to ``youngest brother'' rather than ``youngest player''. BM$\mathcal{X}$ with WQA best matches the intended query, demonstrating the importance of combining lexical similarity with semantic understanding in effective search.

\begin{table*}[htbp]
\setlength\tabcolsep{2.5pt}
\scriptsize
\centering
\begin{threeparttable}
\begin{tabular}{ll}
\toprule
\textbf{Query:} & Who is the youngest player in the Premier League? \\
\toprule
\multicolumn{2}{c}{\textit{BM25}} \\
\midrule
\multirow{3}{*}{\#1} & Thulani Ngcepe (born 19 January 1990 in Tsakane, Gauteng) is a South African football (soccer) striker for Supersport \\
& United in the \textcolor{deeporange}{Premier} Soccer \textcolor{deeporange}{League}. On 14 January 2007,  he became the \textcolor{deeporange}{youngest player ever} to score a goal in the \\
& SA \textcolor{deeporange}{Premier League} after scoring on his debut against Silver Stars. \\
\midrule

\multirow{2}{*}{\#2} & This is a list of \textcolor{deeporange}{Premier League} players who have made 300 or more appearances in the Premier League. Statistics are \\
& updated as of 24 May 2015. Current Premier League players are shown in bold. \\

\midrule

\multirow{3}{*}{\#3} & Ismail Youssef Awadallah Mohamed (born 28 June 1964 in Giza) is a former Egyptian football player. He is currently \\
& the assistant manager of the Egyptian \textcolor{deeporange}{Premier League} giants; Zamalek. Ismail is the \textcolor{deeporange}{youngest} brother of two football \\ 
& players; El Sayed and Ibrahim. \\

\midrule
\midrule
\multicolumn{2}{c}{\textit{BM$\mathcal{X}$}} \\
\midrule

\multirow{3}{*}{\#1} & Thulani Ngcepe (born 19 January 1990 in Tsakane, Gauteng) is a South African football (soccer) striker for Supersport \\
& United in the \textcolor{deeporange}{Premier} Soccer \textcolor{deeporange}{League}. On 14 January 2007,  he became the \textcolor{deeporange}{youngest player ever} to score a goal in the \\
& SA \textcolor{deeporange}{Premier League} after scoring on his debut against Silver Stars. \\
\midrule

\multirow{3}{*}{\#2} & Ismail Youssef Awadallah Mohamed (born 28 June 1964 in Giza) is a former Egyptian football player. He is currently \\
& the assistant manager of the Egyptian \textcolor{deeporange}{Premier League} giants; Zamalek. Ismail is the \textcolor{deeporange}{youngest} brother of two football \\ 
& players; El Sayed and Ibrahim. \\
\midrule 

\multirow{3}{*}{\#3} & Yevgeni Kirillovich Zezin (born April 14, 1976) is a retired Russian professional football player. He was the \textcolor{deeporange}{youngest} \\
& \textcolor{deeporange}{player to ever} play in the Russian \textcolor{deeporange}{Premier League} (he played his first game in 1992 when he was 16 years 114 days \\
& old) until Aleksei Rebko has beat his record in 2002. \\

\midrule
\midrule
\multicolumn{2}{c}{\textit{BM$\mathcal{X}$ + WQA}} \\
\midrule

\multirow{3}{*}{\#1} & Thulani Ngcepe (born 19 January 1990 in Tsakane, Gauteng) is a South African football (soccer) striker for Supersport \\
& United in the \textcolor{deeporange}{Premier} Soccer \textcolor{deeporange}{League}. On 14 January 2007,  he became the \textcolor{deeporange}{youngest player ever} to score a goal in the \\
& SA \textcolor{deeporange}{Premier League} after scoring on his debut against Silver Stars. \\
\midrule

\multirow{3}{*}{\#2} & Scott Robinson (born 12 March 1992) is a Scottish professional footballer who plays for Kilmarnock in the Scottish \\
& Premiership, as a midfielder. He previously played for Heart of Midlothian, where he made his debut aged 16, \\
& becoming the \textcolor{deeporange}{youngest ever player} to appear in the Scottish \textcolor{deeporange}{Premier League} (SPL). \\
\midrule

\multirow{3}{*}{\#3} & Neil Finn (born 29 December 1978 in Rainham, Essex) is a former professional footballer. He played one game for \\
& West Ham United in the \textcolor{deeporange}{Premier League} during the 1995-96 season, aged 17, becoming at that time the \textcolor{deeporange}{youngest} \\
& \textcolor{deeporange}{player to ever} to appear in a Premiership match. \\

\bottomrule
\end{tabular}
\end{threeparttable}
\caption{The top 3 retrieved documents for a given query by different lexical search models on the DBPedia dataset from the BEIR benchmark. The query intents in the retrieved documents are highlighted in dark orange.}
\label{table::case}
\end{table*}

\subsection{Ablation Study}
\label{sec::ablation}
Our ablation study on the SciFact dataset examines the impact of the hyperparameters $\alpha$ and $\beta$ on BM$\mathcal{X}$'s performance (Table \ref{table::ablation}). 
Results vary from $65.22$ to $69.11$, indicating the importance of parameter tuning. 
The result of the default dynamic parameters is $69.42$, 
outperforming all results in Table \ref{table::ablation}. 
While effective for English, these dynamic parameters may require adjustment for other languages or domains.

\begin{table*}[htbp]
\setlength\tabcolsep{2.5pt}
\small
\centering
\begin{threeparttable}
\begin{tabular}{l|ccccccccccc}
\toprule
Params & $\alpha=0.1$ & $\alpha=0.2$ & $\alpha=0.3$ & $\alpha=0.4$ & $\alpha=0.5$ & $\alpha=0.6$ & $\alpha=0.7$ & $\alpha=0.8$ & $\alpha=0.9$ & $\alpha=1.0$ \\
\toprule
$\beta=0.1$ & $65.92$ & $67.13$ & $67.37$ & $67.96$ & $68.25$ & $68.47$ & $68.57$ & $68.47$ & $68.76$ & $68.91$\\
$\beta=0.2$ & $66.00$ & $67.23$ & $67.43$ & $67.94$ & $68.28$ & $68.38$ & $68.66$ & $68.37$ & $68.73$ & $69.02$\\
$\beta=0.3$ & $65.98$ & $66.95$ & $67.44$ & $67.81$ & $68.17$ & $68.56$ & $68.83$ & $68.40$ & $68.55$ & $69.00$\\
$\beta=0.4$ & $65.87$ & $67.01$ & $67.39$ & $67.37$ & $68.17$ & $68.49$ & $68.65$ & $68.49$ & $68.45$ & $69.11$\\
$\beta=0.5$ & $65.78$ & $67.00$ & $67.38$ & $67.35$ & $68.11$ & $68.41$ & $68.60$ & $68.35$ & $68.38$ & $69.11$\\
$\beta=0.6$ & $65.77$ & $66.99$ & $67.14$ & $67.30$ & $68.13$ & $68.25$ & $68.50$ & $68.43$ & $68.50$ & $68.86$\\
$\beta=0.7$ & $65.60$ & $66.74$ & $67.08$ & $67.32$ & $67.80$ & $68.26$ & $68.48$ & $68.35$ & $68.46$ & $68.85$\\
$\beta=0.8$ & $65.47$ & $66.58$ & $66.95$ & $67.29$ & $67.69$ & $68.26$ & $68.37$ & $68.18$ & $68.45$ & $68.76$\\
$\beta=0.9$ & $65.35$ & $66.53$ & $66.92$ & $67.34$ & $67.50$ & $68.23$ & $68.25$ & $68.08$ & $68.36$ & $68.75$\\
$\beta=1.0$ & $65.22$ & $66.24$ & $66.87$ & $67.30$ & $67.41$ & $68.21$ & $68.19$ & $68.12$ & $68.37$ & $68.36$\\

\bottomrule
\end{tabular}
\end{threeparttable}
\caption{BM$\mathcal{X}$ results of the SciFact dataset from the BEIR benchmark with different hyperparameters $\alpha$ and $\beta$. NDCG@10 is the reported metric.}
\label{table::ablation}
\end{table*}

\section{Conclusion}
\label{sec::conclusion}

In this paper, we revisited BM25 and introduced a lexical search algorithm, called BM$\mathcal{X}$. Our approach integrates the proposed entropy-weighted similarity metric with TF-IDF, creating an overall more robust lexical search algorithm. By adding the similarity between queries and documents, BM$\mathcal{X}$ demonstrates a better ability to retrieve relevant documents. Moreover, we developed a weighted query augmentation technique that introduces semantic understanding into lexical search, further improving its effectiveness. This helps us bridge the gap between pure lexical matching and semantic comprehension, addressing a longstanding challenge in information retrieval. We introduced Baguetter, an evaluation framework for IR with a reference implementation of BM$\mathcal{X}$. Additionally, we present a normalization technique for both BM$\mathcal{X}$ and BM25 scores. Our experiments across different information retrieval benchmarks provide evidence of BM$\mathcal{X}$'s ability to improve lexical search performance. 

\section{Acknowledgement}
\label{sec::acknowledgement}
We'd like to thank Darius Koenig for his thorough proofreading.

\bibliography{colm2024_conference}
\bibliographystyle{colm2024_conference}

\appendix
\section{Prompt for Query Augmentation}
\label{sec::query-aug-prompt}
For the query augmentation, we use GPT4 with the prompt in Table \ref{table::prompt}.

\begin{table}[thbp]
    \centering
    \caption{Prompt used in GPT4 for weighted query augmentation.}

    \begin{tabular}{p{0.8\linewidth} }
    \toprule
\small{\fontfamily{ppl}\selectfont

You are an intelligent query augmentation tool. Your task is to augment each query with \{size\} similar queries and output JSONL format, like \{"query": "original query", "augmented\_queries": ["augmented query 1", "augmented query 2", ...]\} \newline

Input query: \{query\} \newline

Output:
}\\
\bottomrule
\end{tabular}
\label{table::prompt}
\end{table}

\section{Enhanced BM25}
\label{sec::enhanced-bm25}
Here, we describe the weighted query augmentation for BM25 and normalized BM25.

\subsection{Weighted Query Augmentation for BM25}
The weighted query augmentation for BM25 is similar to that for BM$\mathcal{X}$.
First, we augment the query using LLMs as follows:
\begin{equation}
\mathcal{Q}^A = \mathrm{LLM}(Q, t),
\end{equation}
where $t$ is the number of augmented queries, and $\mathcal{Q}^A$ denotes the set of augmented queries $\{Q_1^A, Q_2^A, \ldots, Q_t^A\}$.
Then, the weighted query augmentation for BM25 is as follows:
\begin{equation}
\mathrm{score}(D, Q, \mathcal{Q}^A) = \mathrm{score}(D, Q) + \sum_{i=1}^t w_i \cdot \mathrm{score}(D, Q_i^A),
\end{equation}
where $w_i$ is the weight for the $i$-th augmented query, typically set between $0$ and $1$.

\subsection{Normalized BM25}
Similar to BM$\mathcal{X}$, we first estimate the maximum value of the BM25 score:
\begin{equation}
    \mathrm{score}(D, Q)^{max} = \sum_i^m \mathrm{max}(\mathrm{IDF}(q_i)) = m \cdot \left [ \mathrm{log} \left (1 + \frac{ n - 0.5}{1.5} \right ) \right ]
\end{equation}
The normalized BM25 score is then calculated as follows:
\begin{equation}
    \mathrm{score}(D, Q)^{norm} = \frac{\mathrm{score}(D, Q)}{\mathrm{score}(D, Q)^{max}}.
\end{equation}

\end{document}